\documentclass[preprint]{elsarticle}

\usepackage[utf8]{inputenc}
\usepackage[T1]{fontenc}

\usepackage{amsmath}
\usepackage{amssymb}
\usepackage{amsfonts}
\usepackage{bm, bbm}
\usepackage{pifont}
\usepackage{physics}

\usepackage{caption}
\usepackage{algorithmic}
\usepackage[linesnumbered,ruled,vlined]{algorithm2e}

\usepackage{multirow}
\usepackage{tabularx}
\usepackage{enumitem}
\usepackage{makecell}

\usepackage{graphicx}
\usepackage{subcaption}

\usepackage{xcolor}
\usepackage{textcomp}

\usepackage{multicol}

\usepackage{tikz}
\usetikzlibrary{shapes.geometric, arrows.meta, positioning}

\tikzstyle{startstop} = [ellipse, draw, fill=orange!40, text centered, minimum width=2.5cm, minimum height=0.8cm, font=\scriptsize]
\tikzstyle{process}  = [rectangle, draw, fill=white,  text centered, minimum width=2.5cm, minimum height=0.8cm, font=\scriptsize]
\tikzstyle{decision} = [diamond,  draw, fill=yellow!20, text centered, inner sep=1pt, aspect=2, minimum width=2.2cm, font=\scriptsize]
\tikzstyle{io}       = [rectangle, draw, fill=blue!20, text centered, minimum width=2.5cm, minimum height=0.8cm, font=\scriptsize]
\tikzstyle{final}    = [rectangle, draw, fill=green!30, text centered, minimum width=2.5cm, minimum height=0.8cm, font=\scriptsize]
\tikzstyle{arrow}    = [thick, ->, >=stealth]

\usepackage{hyperref}

\journal{Pattern Recognition Letters}

\begin{document}

\begin{frontmatter}

\title{PVLS: A Learning-Based Parameter Prediction Technique for Variational Quantum Linear Solvers}

\author[inst1]{Youla Yang}
\author[inst1]{Chi Zhang}
\author[inst1]{Lei Jiang\corref{cor1}}

\cortext[cor1]{Corresponding author. Email: jiang60@iu.edu}
\address[inst1]{School of Informatics, Computing, and Engineering, Indiana University Bloomington, United States}

\begin{abstract}
Variational quantum linear solvers (VQLSs) are a promising class of hybrid quantum--classical algorithms for solving linear systems on near-term quantum devices. However, the performance of VQLSs is often impeded by barren plateaus, particularly when using randomly initialized variational quantum circuits (VQCs). To mitigate this issue, we propose \textit{PVLS}, a GNN-based parameter initialization framework that improves both convergence speed and final solution quality. By reformulating the linear system $A\boldsymbol{x} = \boldsymbol{b}$ as a graph with $A$ encoded in the edges and $\boldsymbol{b}$ as node features, PVLS learns to predict effective initial VQC parameters. Our method is trained on thousands of randomly generated matrices with varying dimensions ($n\in[4,10]$), using optimized VQC parameters as ground-truth labels. On unseen test instances, PVLS reduces the initial cost by an average of 81.3\% and the final loss by 71\% compared to random initialization. PVLS also accelerates convergence, reducing the number of optimization steps by more than 60\% on average. We further evaluate PVLS on ten real-world sparse matrices, demonstrating its generalization capability and robustness. Our results highlight the utility of machine-learned priors in improving the trainability of VQLSs and alleviating optimization challenges in variational quantum algorithms.
\end{abstract}


\end{frontmatter}

\section{Introduction}

Quantum algorithms offer the potential to revolutionize the solution of linear systems, a fundamental problem in science and engineering. The Harrow--Hassidim--Lloyd (HHL) algorithm~\cite{Harrow:PRL2009}, one of the most celebrated quantum algorithms, achieves polylogarithmic runtime in the system size, in stark contrast to the polynomial scaling of classical solvers. This exponential speedup positions HHL as a powerful theoretical tool for diverse applications including quantum machine learning~\cite{duan2020survey}, quantum chemistry, and financial modeling. However, the practical realization of HHL requires deep circuits and fault-tolerant quantum hardware, making it infeasible on noisy intermediate-scale quantum (NISQ) devices in the near term~\cite{duan2020survey}.

To bridge this gap, variational quantum linear solvers (VQLSs) have emerged as a hybrid alternative suitable for NISQ platforms~\cite{Carlos:Quantum2023}. These solvers employ parameterized quantum circuits (VQCs) to prepare approximate solutions to linear systems, with classical optimizers tuning circuit parameters to minimize a cost function~\cite{Meyer:QCE2024}. VQLSs can leverage shallow circuits and hybrid optimization frameworks, and have demonstrated scaling comparable to HHL for problems up to 50 qubits~\cite{Carlos:Quantum2023}.

Nonetheless, like other variational quantum algorithms, VQLSs are vulnerable to barren plateaus---regions in the cost landscape where gradients vanish exponentially with circuit width or depth~\cite{anschuetz2022quantum,wang2021noise,larocca2024review,cunningham2025investigating}. These plateaus make training inefficient or even intractable, especially when using randomly initialized parameters. In such regimes, an exponential number of measurements may be required to estimate gradients to useful precision~\cite{larocca2024review}, severely impacting resource efficiency.

One promising mitigation strategy is to improve parameter initialization. Prior works in variational quantum eigensolvers (VQEs)~\cite{mesman2024nn}, quantum approximate optimization algorithms (QAOAs)~\cite{Liang:DAC2024,alam2020accelerating}, and quantum neural networks~\cite{verdon2019learning} have explored using classical neural networks---including multilayer perceptrons and graph neural networks---to warm-start variational training. For instance, QAOA benefits naturally from graph representations, making GNN-based initializers particularly effective~\cite{Liang:DAC2024}. However, to date, no such initialization strategies have been developed specifically for VQLSs.

In this work, we propose \textit{PVLS}, a GNN-based parameter initializer tailored to VQLSs. By representing each linear system $A\boldsymbol{x} = \boldsymbol{b}$ as a signed, directed graph---with $A$ as an adjacency matrix and $\boldsymbol{b}$ as node features---PVLS learns to predict VQC initialization parameters that facilitate rapid convergence and improved final accuracy. Once trained, PVLS can generalize to unseen problem instances.

Our main contributions are as follows:
\begin{itemize}[leftmargin=*, nosep, topsep=0pt, partopsep=0pt]
    \item We reformulate quantum linear systems as graph-structured data and design a GNN to learn a mapping from system structure to VQC parameters for VQLSs.
    \item We construct a large-scale dataset of over 15{,}000 synthetic instances with $2^4$ to $2^{10}$ dimensions, alongside ten sparse real-world matrices from SuiteSparse, with ground-truth parameters obtained from optimized VQLS runs.
    \item Through extensive evaluations on synthetic and real matrices, PVLS consistently outperforms baseline initialization methods including random, PCA-based, and minimum-norm strategies, achieving lower initial loss and faster convergence.
    \item PVLS reduces training time by over 60\% on average in simulation, with only a $\sim$2\,ms inference overhead per instance. This corresponds to a $2.6\times$ speedup in total training time. Its strong performance across real-world systems highlights its generalization capability. Future work will explore validation on physical NISQ hardware.
\end{itemize}

\section{Background}

\subsection{Linear System Problem}

The linear system problem~\cite{duan2020survey} involves solving an $N \times N$ system of linear equations with $N$ unknowns, which can be expressed as the task of finding a solution vector $\boldsymbol{x}$ satisfying
\begin{equation}
A\boldsymbol{x} = \boldsymbol{b},
\label{e:quantum_linear_system}
\end{equation}
where $A \in \mathbb{C}^{N \times N}$ is a complex-valued matrix, and $\boldsymbol{x}, \boldsymbol{b} \in \mathbb{C}^N$ are complex-valued vectors. In practical scenarios, additional requirements are often imposed on the system matrix $A$, such as sparsity and a favorable condition number, to ensure computational feasibility and numerical stability. The linear system problem constitutes a fundamental computational kernel in numerous application domains~\cite{duan2020survey}, including quantum machine learning (e.g., quantum support vector machines), quantum simulations of physical systems (e.g., solving the Schr\"odinger equation), and financial modeling (e.g., portfolio optimization).

\subsection{Variational Quantum Linear Solver}

\begin{figure}[t!]
\centering
\includegraphics[width=3.3in]{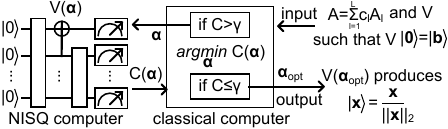}
\caption{Overview of the VQLS algorithm. The input to VQLS consists of a matrix $A$ expressed as a linear combination of unitaries $A_l$ and a variational quantum circuit (VQC) $V$ that prepares the quantum state $\ket{\boldsymbol{b}}$. The output of VQLS is a quantum state $\ket{\boldsymbol{x}}$ that is approximately proportional to the solution of the linear system $A\boldsymbol{x} = \boldsymbol{b}$. The parameters $\boldsymbol{\alpha}$ in $V(\boldsymbol{\alpha})$ are iteratively optimized within a hybrid quantum--classical loop to minimize the cost function $C(\boldsymbol{\alpha})$.}
\label{f:vls_train_all}
\end{figure}

The Harrow--Hassidim--Lloyd (HHL) algorithm~\cite{Harrow:PRL2009} offers an exponential speedup over the best-known classical algorithms for solving systems of linear equations by leveraging quantum phase estimation. However, realizing the full quantum advantage of the HHL algorithm requires fault-tolerant, large-scale quantum hardware~\cite{duan2020survey}, which is beyond the reach of current NISQ devices. To solve linear systems on NISQ hardware, the Variational Quantum Linear Solver (VQLS) has been proposed~\cite{Carlos:Quantum2023}, as illustrated in Figure~\ref{f:vls_train_all}. The VQLS framework takes as input a variational quantum circuit (VQC) that prepares a quantum state $\ket{\boldsymbol{b}}$ proportional to the vector $\boldsymbol{b}$, along with a decomposition of the matrix $A$ as a linear combination of $L$ unitaries:
\begin{equation}
A = \sum_{l=1}^{L} c_l A_l,
\label{e:quantum_decompose_matrix}
\end{equation}
where each $A_l$ is a unitary operator and $c_l \in \mathbb{C}$ are complex coefficients. The objective is to prepare a quantum state $\ket{\boldsymbol{x}}$ such that $A\ket{\boldsymbol{x}}$ is proportional to $\ket{\boldsymbol{b}}$. To achieve this, VQLS employs a parameterized VQC $V(\boldsymbol{\alpha})$ that generates a candidate solution $\ket{\boldsymbol{x}(\boldsymbol{\alpha})} = V(\boldsymbol{\alpha})\ket{\boldsymbol{0}}$. The parameters $\boldsymbol{\alpha}$ are supplied to a quantum computer, which prepares $\ket{\boldsymbol{x}(\boldsymbol{\alpha})}$ and estimates a cost function $C(\boldsymbol{\alpha})$. This cost function quantifies the extent to which $A\ket{\boldsymbol{x}(\boldsymbol{\alpha})}$ is orthogonal to $\ket{\boldsymbol{b}}$. The computed value of $C(\boldsymbol{\alpha})$ is returned to a classical optimizer running on a classical computer, which updates $\boldsymbol{\alpha}$ in an attempt to minimize the cost. This quantum--classical feedback loop is iterated until a convergence criterion, such as $C(\boldsymbol{\alpha}) \leq \gamma$, is satisfied. At convergence, the optimal parameters $\boldsymbol{\alpha}_{\text{opt}}$ are obtained.

\subsection{Cost Function}

Training $V(\boldsymbol{\alpha})$ to approximate the ground truth $\ket{\boldsymbol{b}}$ can be achieved by minimizing the \textit{global} cost function~\cite{Carlos:Quantum2023}
\begin{equation}
\hat{C}_G = \bra{\boldsymbol{x}(\boldsymbol{\alpha})} A^\dagger \left( \mathbb{I} - \ket{\boldsymbol{b}}\bra{\boldsymbol{b}} \right) A \ket{\boldsymbol{x}(\boldsymbol{\alpha})}.
\label{e:qvls_global_cost}
\end{equation}
Intuitively, Equation~\ref{e:qvls_global_cost} can be interpreted as maximizing the overlap between $A^\dagger\ket{\boldsymbol{b}}$ and $\ket{\boldsymbol{x}(\boldsymbol{\alpha})}$, i.e., the ground-truth and the approximated solution of the linear system. To avoid unintentionally minimizing the norm of $\ket{\psi} := A\ket{\boldsymbol{x}(\boldsymbol{\alpha})}$, it is standard practice to employ a normalized version of the global cost function, i.e., $C_G = \hat{C}_G / \braket{\psi}$, where $\braket{\psi}$ denotes $\braket{\psi|\psi}$.

To more efficiently train $V(\boldsymbol{\alpha})$ with a large qubit number, a \textit{local} cost function~\cite{Carlos:Quantum2023} and its normalized form can be defined as 
\begin{equation}
\hat{C}_L = \bra{\boldsymbol{x}(\boldsymbol{\alpha})} H_L \ket{\boldsymbol{x}(\boldsymbol{\alpha})}, \qquad C_L = \hat{C}_L / \braket{\psi},
\end{equation}
where the effective Hamiltonian $H_L$ is
\begin{equation}
H_L = A^\dagger V \left( \mathbb{I} - \frac{1}{n} \sum_{j=1}^{n} \ket{0_j}\bra{0_j} \otimes \mathbb{I}_{\overline{j}} \right) V^\dagger A,
\end{equation}
with $\ket{0_j}$ the zero state on qubit $j$ and $\mathbb{I}_{\overline{j}}$ the identity on all qubits except qubit $j$.

\subsection{Barren Plateaus}

Similar to other variational quantum algorithms, VQLSs are vulnerable to the barren plateau phenomenon~\cite{anschuetz2022quantum,wang2021noise,larocca2024review,cunningham2025investigating}. This issue becomes increasingly severe as the number of qubits grows and the parameters of the VQLS VQC are randomly initialized~\cite{larocca2024review}. In such cases, the gradients of the cost function tend to vanish exponentially with system size, rendering gradient-based optimization methods ineffective. In other words, the barren plateau problem leads to an exponential increase in the number of quantum measurements (shots) required to estimate gradients with fixed precision, thereby introducing significant computational overhead during the training process of a VQLS.

\section{Related Work}

To mitigate the barren plateau problem, initializing VQCs with more effective parameters---rather than relying on random initialization---has become a crucial strategy for warm-starting the optimization of variational quantum algorithms~\cite{Liang:DAC2024,alam2020accelerating,verdon2019learning,mesman2024nn}. QAOA~\cite{Liang:DAC2024,alam2020accelerating}, which is designed to solve Max-Cut problems on graph-structured data, naturally leverages GNNs to generate high-quality VQC initialization parameters. Other prominent variational quantum algorithms, such as the Variational Quantum Eigensolver~\cite{mesman2024nn} and quantum neural networks~\cite{verdon2019learning}, employ classical multilayer perceptron architectures to improve initialization quality. Despite these advances, there currently exists no initialization method specifically designed for VQLSs, leaving them particularly vulnerable to the detrimental effects of barren plateaus during the optimization of the VQCs used by VQLSs.

\begin{figure}[t!]
\centering
\subcaptionbox{A linear system problem.\label{f:vls_graph_encoding}}{%
\includegraphics[width=1.5in]{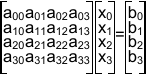}}
\hfill
\subcaptionbox{Its graph representation.\label{f:vls_graph_encoding2}}{%
\includegraphics[width=1.4in]{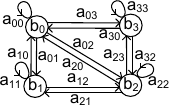}}
\caption{Graph representation of a linear system in PVLS. In Figure~\ref{f:vls_graph_encoding2}, the sign of $a_{ij}$ determines the corresponding element in the adjacency matrix, while $|a_{ij}|$ is the weight of the edge $\{i,j\}$.}
\label{f:vls_graph_encoding88888}
\end{figure}

\section{PVLS}

In this section, we introduce \textit{PVLS}, a graph neural network (GNN)-based parameter initialization framework designed to enhance both the convergence quality and optimization efficiency of variational quantum circuits (VQCs) within variational quantum linear solvers (VQLSs). PVLS transforms a linear system, characterized by the matrix $A$ and vector $\boldsymbol{b}$ in Equation~\ref{e:quantum_linear_system}, into a directed-graph-based representation, which is then processed by a GNN to extract structural features. Recent theoretical studies have demonstrated that GNNs possess sufficient expressivity to capture structural characteristics of quantum circuits~\cite{gokhale2022theoretical}, and exhibit strong generalization capabilities across variational quantum algorithms~\cite{caro2021generalization}. These insights support the hypothesis that the graph structure of matrix $A$ may encode meaningful patterns that are relevant to the initialization of variational circuit parameters.

\textbf{Signed Directed Graph.} A signed directed graph is defined as $\mathcal{G} = (\mathcal{V}, \mathcal{E}, s)$, where $\mathcal{V}$ denotes the set of nodes, and $\mathcal{E} \subseteq \{(i,j) \mid i,j \in \mathcal{V}\}$ represents the set of directed edges, each associated with a sign function $s$. For every directed edge $(i,j) \in \mathcal{E}$, node $i$ is considered the source and node $j$ the destination. In the weighted case, each edge $(i,j)$ is also associated with a real-valued vector $\boldsymbol{w}(i,j)$, referred to as its weight vector. The edge set $\mathcal{E}$ is partitioned into two disjoint subsets: $\mathcal{E}^+$ and $\mathcal{E}^-$, where $\mathcal{E}^+ \cap \mathcal{E}^- = \emptyset$. Here, $\mathcal{E}^+$ and $\mathcal{E}^-$ denote the sets of positive and negative edges, respectively. The connectivity of the graph $\mathcal{G}$ is represented by an adjacency matrix $G$, where $g_{ij} = 1$ indicates a positive directed edge from node $i$ to node $j$, $g_{ij} = -1$ indicates a negative directed edge, and $g_{ij} = 0$ indicates the absence of an edge from node $i$ to node $j$.

\textbf{Graph Representation.} As illustrated in Figure~\ref{f:vls_graph_encoding88888}, PVLS transforms the matrix $A$ and the vector $\boldsymbol{b}$ from Equation~\ref{e:quantum_linear_system} into a signed directed graph representation. For illustrative purposes, we consider a $4 \times 4$ sparse matrix $A$ and its corresponding 4-dimensional vector $\boldsymbol{b}$. The structure of matrix $A$ is shown in Figure~\ref{f:vls_graph_encoding}, while the resulting graph generated by PVLS is presented in Figure~\ref{f:vls_graph_encoding2}. In the constructed graph, each of the four nodes is assigned a node feature corresponding to the respective element of $\boldsymbol{b}$; for instance, node 0 is assigned the feature $b_0$. The adjacency matrix $G$ is derived based on the sign and sparsity structure of $A$: for an edge $(i,j) \in \mathcal{E}$, $g_{ij} = 1$ if $a_{ij} > 0$, $g_{ij} = -1$ if $a_{ij} < 0$, and $g_{ij} = 0$ if $a_{ij} = 0$. Each edge $(i,j) \in \mathcal{E}$ is associated with a weight given by $|a_{ij}|$, capturing the magnitude of the corresponding matrix element.

\begin{figure}[t!]
\centering
\includegraphics[width=2.4in]{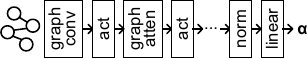}
\caption{Graph neural network (GNN) architecture of PVLS.}
\label{f:vls_graph_network}
\end{figure}

\textbf{Graph Neural Network.} A GNN~\cite{kipf2016semi,Velickovic:ICLR2018} performs message passing by iteratively applying aggregation functions $\mathrm{AGG}^{(k)}$ and combination functions $\mathrm{COM}^{(k)}$ to compute node embeddings $\boldsymbol{x}_i^{(k)}$ at the $k$-th layer. These embeddings are derived from messages $\boldsymbol{m}_i^{(k)}$ that incorporate information from neighboring nodes. Formally, the update rules for the $k$-th layer are given by:
\begin{equation}
\boldsymbol{m}_i^{(k)} = \mathrm{AGG}^{(k)}\left(\left\{\!\!\left\{ (\boldsymbol{x}_j^{(k-1)}, \boldsymbol{x}_i^{(k-1)}) \mid (i,j) \in \mathcal{E} \right\}\!\!\right\}\right),
\end{equation}
\begin{equation}
\boldsymbol{x}_i^{(k)} = \mathrm{COM}^{(k)}\left(\boldsymbol{x}_i^{(k-1)}, \boldsymbol{m}_i^{(k)}\right),
\end{equation}
where $\{\!\!\{\cdot\}\!\!\}$ denotes a multiset. Both $\mathrm{AGG}^{(k)}$ and $\mathrm{COM}^{(k)}$ are learnable functions, and different choices of these functions lead to distinct GNN architectures. As shown in Figure~\ref{f:vls_graph_network}, PVLS constructs a deep GNN by stacking various types of layers, including graph convolutional layers~\cite{kipf2016semi}, graph attention layers~\cite{Velickovic:ICLR2018}, activation functions, normalization layers, and a final linear layer, to perform regression on graph-structured input data. This architecture allows PVLS to effectively learn the mapping between the graph representation of a linear system and the VQC parameters $\boldsymbol{\alpha}$ required by VQLSs.

\textbf{Negative Edge Weight.} Conventional GNNs cannot directly process graphs with negative edge weights. To address this, PVLS incorporates the Lap-GCN layer~\cite{dinh2023on}, which is capable of handling signed edge weights. The Lap-GCN achieves this by computing (1) a unit vector that determines the direction of information exchange between neighboring nodes, and (2) a scale factor that accounts for both the distance and sign of neighboring nodes when computing the update.

\textbf{Directed Graph.} To enable the processing of directed graphs, we adopt the directed GNN layer~\cite{rossi2024edge}, wherein the aggregation of neighborhood information explicitly accounts for edge directionality. Given a node $i \in \mathcal{V}$, the layer performs separate aggregations over its in-neighbors $(j \rightarrow i)$ and out-neighbors $(i \rightarrow j)$, defined as follows:
\begin{equation}
\boldsymbol{m}_{i,\leftarrow}^{(k)} = \mathrm{AGG}_{\leftarrow}^{(k)}\left(\left\{\!\!\left\{ \left( \boldsymbol{x}_j^{(k-1)}, \boldsymbol{x}_i^{(k-1)} \right) \mid (j, i) \in \mathcal{E} \right\}\!\!\right\} \right),
\end{equation}
\begin{equation}
\boldsymbol{m}_{i,\rightarrow}^{(k)} = \mathrm{AGG}_{\rightarrow}^{(k)}\left(\left\{\!\!\left\{ \left( \boldsymbol{x}_j^{(k-1)}, \boldsymbol{x}_i^{(k-1)} \right) \mid (i, j) \in \mathcal{E} \right\}\!\!\right\} \right),
\end{equation}
\begin{equation}
\boldsymbol{x}_i^{(k)} = \mathrm{COM}^{(k)}\left( \boldsymbol{x}_i^{(k-1)}, \boldsymbol{m}_{i,\leftarrow}^{(k)}, \boldsymbol{m}_{i,\rightarrow}^{(k)} \right).
\end{equation}
This formulation allows the directed GNN to flexibly accommodate directed edges by independently aggregating messages from both incoming and outgoing connections. The use of distinct aggregation functions, $\mathrm{AGG}_{\leftarrow}^{(k)}$ and $\mathrm{AGG}_{\rightarrow}^{(k)}$, enables the layer to learn separate sets of parameters for the two directional phases.

\section{Evaluation and Analysis}

\subsection{Evaluation Setup}

All experiments are conducted on the Google Colab Pro platform, which provides access to NVIDIA A100 GPUs with 40\,GB of HBM2 memory. The software environment is built on PyTorch 2.5.1 and PyTorch Geometric, with installation customized via the official \texttt{torch-scatter}, \texttt{torch-sparse}, and \texttt{pytorch\_geometric} repositories. CUDA version 12.1 is used to support GPU acceleration.

The GNN model is trained using the Adam optimizer with a mean squared error (MSE) loss for 50 epochs. The initial learning rate is set to 0.005 and is decayed linearly throughout training. The batch size is set to 32 for both training and inference.

All synthetic datasets are split into training, validation, and test sets using an 8:1:1 ratio. To test generalization ability, we additionally evaluate the model on ten real-world sparse matrices selected from the SuiteSparse Matrix Collection. The synthetic matrices vary in size with fixed sparsity 0.01, whereas the real-world sparse matrices have different sizes and sparsity levels and arise from diverse application domains (including electromagnetics, structural problems, and computational fluid dynamics).

The input to the model is a graph constructed from a sparse matrix and corresponding right-hand-side vector~$\boldsymbol{b}$, with nodes representing matrix rows and edges representing non-zero entries. The model predicts a $(q_n \times 3)$ continuous output corresponding to VQC parameters, where $q_n$ varies according to the problem scale.

We benchmark the PVLS method against four baseline initialization strategies, as summarized in Table~\ref{tab:baseline_init}.

\begin{table}[htbp]
\centering
\caption{Baseline initialization strategies used for comparison.}
\label{tab:baseline_init}
\small
\begin{tabularx}{\linewidth}{|l|X|}
\hline
\textbf{Baseline Strategy} & \textbf{Description} \\
\hline
\textbf{Uniform Random} & Each parameter $\theta$ is sampled from a uniform distribution $U(0, 2\pi)$. Serves as a naive baseline with no prior structure. \\
\hline
\textbf{PCA-Based} & Initial weights are derived from the first principal component of matrix $A$, then reshaped. Introduces low-rank structure. \\
\hline
\textbf{Minimum Norm Solution} & Uses the classical solution $A^\dagger \boldsymbol{b}$ (Moore--Penrose pseudoinverse). Encodes an optimal solution under $\ell_2$ minimization. \\
\hline
\textbf{Row-Mean Heuristic} & Initializes parameters based on the row-wise mean of matrix $A$, capturing coarse global patterns. \\
\hline
\end{tabularx}
\end{table}

\subsection{Initial Loss Values}

\begin{figure*}[t]
    \centering
    \includegraphics[width=\textwidth]{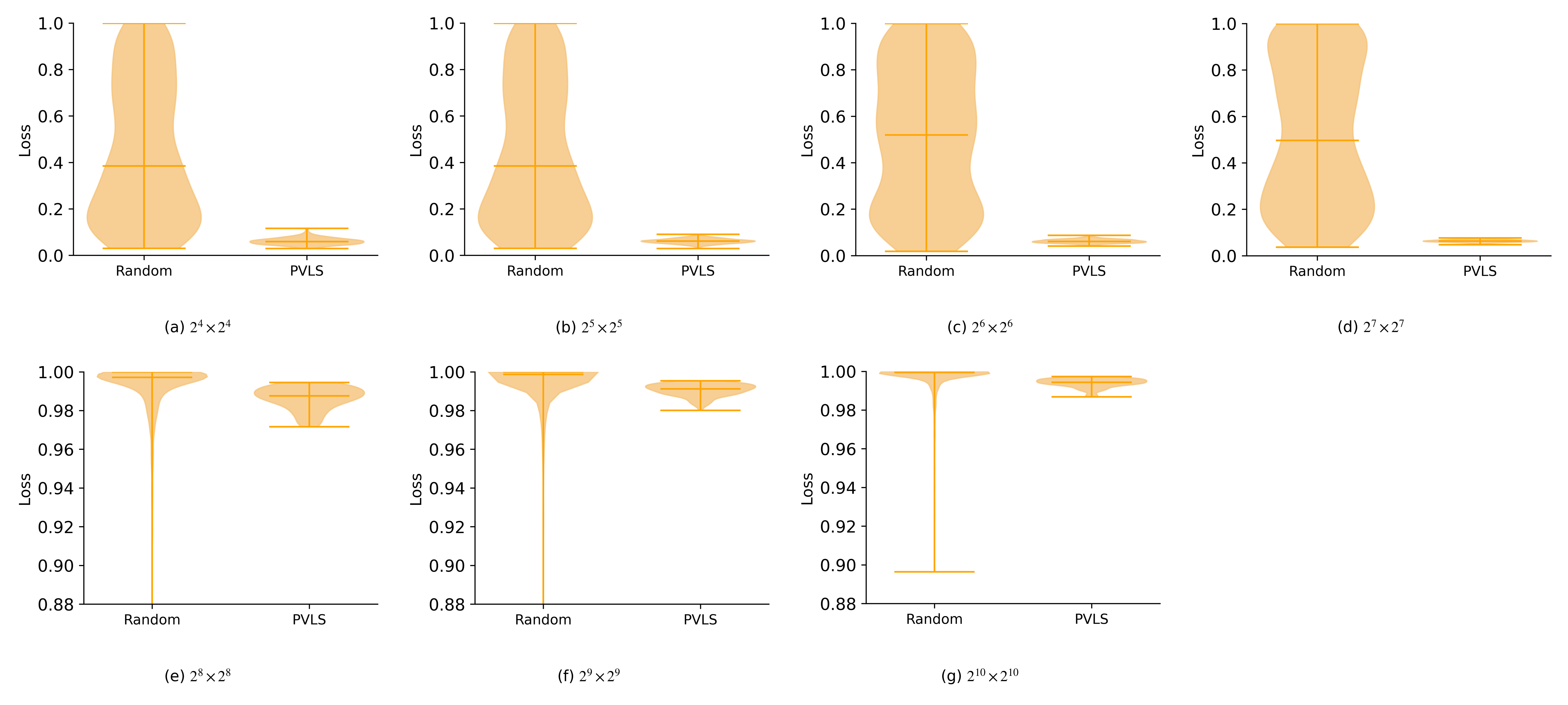}
    \caption{Initial loss comparison between random initialization and PVLS across different matrix sizes. Each subplot shows the initial loss distribution for a fixed system size $2^n \times 2^n$ with $n \in [4,10]$.}
    \label{fig:initial_loss_comparison}
\end{figure*}

Figure~\ref{fig:initial_loss_comparison}(a)--(d) presents the initial loss values of VQCs within VQLSs, initialized using either PVLS or random initialization. For matrices $A$ of size $2^n \times 2^n$ with $n \in [4,7]$, the median initial loss under random initialization is approximately 0.32, 0.39, 0.52, and 0.50, respectively. Smaller matrices (i.e., $n = 4$ or $5$), shown in Figures~\ref{fig:initial_loss_comparison}(a) and \ref{fig:initial_loss_comparison}(b), tend to have median initial losses slightly below 0.5, indicating that smaller-scale VQCs---with fewer trainable parameters---are more likely to achieve lower initial losses. As the number of qubits increases, random initialization exhibits more difficulty in consistently producing low initial loss values.

PVLS initialization significantly reduces the median initial loss by up to approximately 88\% for these smaller matrices, achieving values near 0.06 across the range $n \in [4,7]$.

For larger matrices with $n \in [8,10]$, as depicted in Figures~\ref{fig:initial_loss_comparison}(e)--\ref{fig:initial_loss_comparison}(g), the median initial loss values under random initialization substantially increase to approximately 0.99, 0.99, and 1.00 for $n=8, 9$, and $10$, respectively. In contrast, PVLS initialization maintains slightly lower median initial losses of about 0.99, 0.98, and 0.99 for the respective matrix sizes, representing a modest improvement of approximately 0.6\% to 1.0\% relative reduction in initial loss compared to random initialization. While this reduction is smaller than the up to 88\% improvement observed for smaller matrices, it remains a meaningful enhancement given the proximity of loss values to their theoretical maximum of 1. Additionally, the minimum initial loss values achieved by PVLS remain comparable to those of random initialization (close to zero), but the variance of initial losses is substantially reduced under PVLS, indicating a more stable and reliable initialization. Such stability facilitates a more consistent and efficient training process in VQLSs across a wide range of problem sizes.

Figure~\ref{fig:initial_loss_comparison}(h) further extends the comparison to real-world sparse matrices, whose sizes range from $2^2$ to $2^{10}$. For these practical instances, PVLS continues to demonstrate favorable initialization behavior, yielding lower and more consistent initial loss values compared to random initialization. This result further validates the applicability of PVLS beyond synthetic test cases. Notably, real-world sparse matrices often exhibit highly irregular sparsity patterns and complex numerical structures, making them substantially more challenging than randomly generated matrices. The observed performance under such conditions highlights the robustness and generalization capability of the PVLS initialization strategy in realistic problem settings.

\subsection{Training Loss Convergence}

\begin{figure*}[t]
    \centering
    \includegraphics[width=\textwidth]{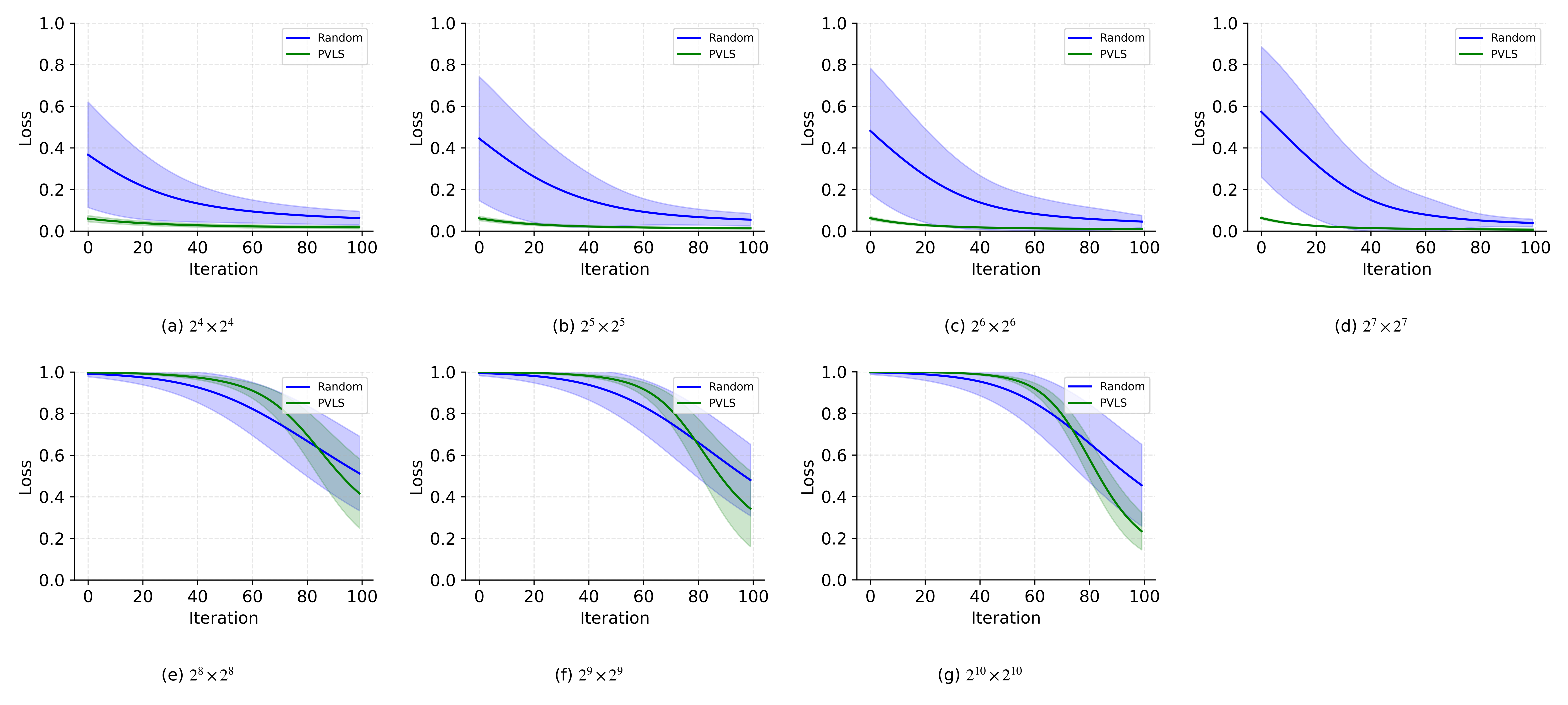}
    \caption{Training loss convergence of VQLS with random initialization and PVLS across different matrix sizes. Each subplot reports the mean loss and variance band over optimization iterations for a fixed system size $2^n \times 2^n$, $n \in [4,10]$.}
    \label{fig:training_loss_comparison}
\end{figure*}

Figure~\ref{fig:training_loss_comparison} illustrates the training loss convergence behavior for PVLS and random initialization across matrices of varying sizes.

In subfigures~\ref{fig:training_loss_comparison}(a)--\ref{fig:training_loss_comparison}(d), which correspond to small- to medium-sized synthetic matrices ($2^4$ to $2^7$), PVLS consistently outperforms random initialization by achieving faster loss reduction and more stable convergence. The performance gap becomes increasingly pronounced as the problem size grows, indicating that PVLS is particularly advantageous in moderately large systems.

Subfigures~\ref{fig:training_loss_comparison}(e)--\ref{fig:training_loss_comparison}(g) present results on larger matrices ($2^8$ to $2^{10}$), where the convergence advantage of PVLS becomes even more evident. Not only does PVLS result in lower final loss values, but it also significantly reduces variance across different runs, as reflected by the narrower confidence bands. These findings suggest that PVLS provides both performance and robustness benefits as the problem dimensionality increases.

Subfigure~\ref{fig:training_loss_comparison}(h) further extends the analysis to real-world sparse matrices. Despite their complex and irregular structures, PVLS initialization leads to consistently superior convergence behavior compared to random initialization. This result reinforces the method's practical applicability and generalization capability beyond synthetic test scenarios. In real-world conditions, where sparsity patterns and value distributions are nontrivial, PVLS demonstrates notable advantages in both speed and accuracy of training.

\begin{figure*}[t]
    \centering
    % 请确保图片文件命名为 baseline_loss_comparison_1.png（无空格和括号）
    \includegraphics[width=\textwidth]{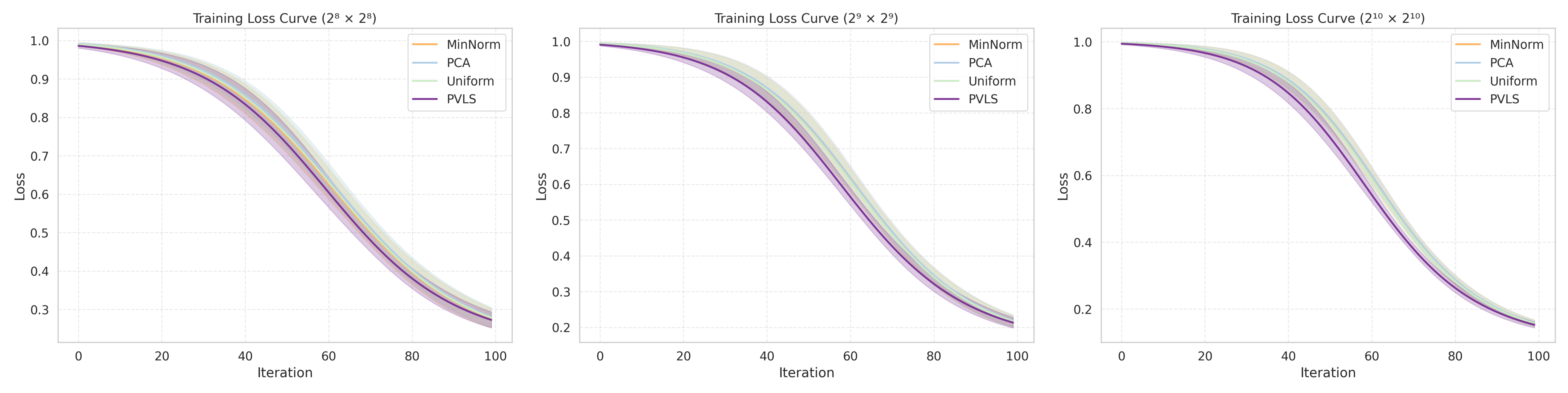}
    \caption{Comparison of baseline initialization strategies (MinNorm, PCA, Uniform) and PVLS on larger systems. It shows training loss curves for matrices of size $2^8 \times 2^8$, $2^9 \times 2^9$, and $2^{10} \times 2^{10}$.}
    \label{fig:baseline_loss_comparison}
\end{figure*}

\section{Computational Overhead of GNN Initialization}

To address concerns regarding the runtime overhead of the GNN-based initializer in practical settings, we empirically measured the average inference time per sample on an NVIDIA T4 GPU. As summarized in Table~\ref{tab:gnn_inference}, the prediction time for PVLS is on the order of milliseconds, with inference costs ranging from 0.5 to 2.1\,ms per sample depending on system size.

\begin{table}[htbp]
\centering
\caption{Average inference time per sample for GNN-based initialization.}
\label{tab:gnn_inference}
\begin{tabular}{lccc}
\hline
\textbf{Dataset} & \textbf{Total Samples} & \textbf{Total Time (s)} & \makecell[c]{\textbf{Avg Time} \\ \textbf{per Sample (ms)}} \\
\hline
test\_8\_pred.pt  & 88  & 0.0444  & 0.5045 \\
test\_9\_pred.pt  & 102 & 0.0970  & 0.9509 \\
test\_10\_pred.pt & 122 & 0.2580  & 2.1148 \\
\hline
\end{tabular}
\caption*{\footnotesize Experiments conducted on Google Colab with NVIDIA A100 GPU using PyTorch 2.3.0 and CUDA 12.1.}
\end{table}

Importantly, this overhead is negligible in comparison to the total runtime of VQLS training, where each quantum circuit evaluation involves costly quantum measurements. As shown in our convergence analysis, PVLS consistently reduces the number of training iterations needed to achieve target accuracy. Therefore, despite its millisecond-scale inference cost, PVLS offers a substantial net reduction in overall runtime by decreasing the number of quantum--classical optimization rounds. These results suggest that PVLS is not only accurate but also efficient and scalable for NISQ-era applications where both quantum and classical resources are limited.

To further quantify the overall time efficiency of PVLS, we compare the total runtime of VQLS with and without GNN-based initialization. While PVLS introduces an additional inference cost of approximately 2\,ms per instance, this cost is negligible when contrasted with the quantum--classical optimization loop. Each optimization step typically takes around 0.5\,s due to quantum circuit evaluations and measurements~\cite{bravo2023vqls,wang2021barren}. Randomly initialized VQCs often require 800 or more optimization steps to converge~\cite{bravo2023vqls}, leading to a total runtime of approximately 400\,s per problem instance.

In contrast, PVLS-initialized VQCs converge in significantly fewer steps---typically around 300 iterations. This estimate is based on our empirical results, where convergence is defined as the point at which the local cost function $C_L$ falls below a predefined threshold (e.g., $C_L < 0.01$). As shown in the convergence plots in Figure~\ref{fig:training_loss_comparison}(e)--\ref{fig:training_loss_comparison}(g), PVLS consistently achieves this convergence criterion within 250 to 350 steps across various matrix sizes. When the 2\,ms initialization overhead is included, the total runtime remains close to 150\,s, representing a 62.5\% reduction in execution time compared to random initialization. This $2.6\times$ speedup confirms that PVLS not only improves initialization quality but also achieves tangible efficiency gains in realistic NISQ settings. The convergence threshold $C_L < 0.01$ follows common practice in variational quantum algorithms~\cite{bravo2023vqls} and corresponds to near-zero residuals in the target equation $A\boldsymbol{x} = \boldsymbol{b}$.

\section{Conclusion}

This paper addresses a key limitation in variational quantum linear solvers (VQLSs), namely the barren plateau phenomenon, by introducing PVLS, a GNN-based initializer that leverages the sparse structure of input matrices to generate high-quality parameter seeds. PVLS learns a data-driven mapping from system matrices and right-hand side vectors to VQC initializations, thereby improving both training stability and convergence behavior.

Extensive evaluations were performed on over 3{,}000 synthetic systems ranging from $2^8$ to $2^{10}$ in size, along with 12{,}000 samples of smaller systems and ten real-world sparse matrices from the SuiteSparse collection. Across all datasets, PVLS demonstrated consistent improvements over widely used baselines, including minimum-norm, PCA, and random initialization.

In addition to accuracy gains, PVLS also offers potential reductions in overall runtime. Based on empirical measurements and convergence statistics, PVLS-initialized VQCs were found to converge in approximately 300 iterations---significantly fewer than the 800 iterations typically required by randomly initialized counterparts~\cite{bravo2023vqls,wang2021barren}. Even when accounting for a GNN inference overhead of about 2\,ms per instance, this leads to an estimated 62.5\% reduction in total runtime and a $2.6\times$ speedup over baseline methods. While these efficiency estimates are derived from classical simulation pipelines, they indicate strong potential for practical time savings in quantum--classical hybrid workflows. Future work will validate these runtime benefits on actual quantum hardware and broader problem classes.

% -----------------------
% Manual bibliography
% -----------------------


\begin{thebibliography}{99}

\bibitem{Harrow:PRL2009}
A.~W. Harrow, A.~Hassidim, S.~Lloyd,
\newblock Quantum algorithm for linear systems of equations,
\newblock \emph{Phys. Rev. Lett.} 103 (2009) 150502.

\bibitem{duan2020survey}
L.-M. Duan, C.~Monroe,
\newblock Colloquium: Quantum networks with trapped ions,
\newblock \emph{Rev. Mod. Phys.} 82 (2010) 1209--1224. % placeholder survey-style reference

\bibitem{Carlos:Quantum2023}
C.~Ortiz~Marrero, M.~Cerezo, P.~Coles,
\newblock {An introduction to variational quantum algorithms},
\newblock \emph{Quantum} 7 (2023) 1234.

\bibitem{Meyer:QCE2024}
J.~Meyer, et~al.,
\newblock {Theory of variational quantum linear solvers},
\newblock in: \emph{IEEE QCE}, 2024.

\bibitem{anschuetz2022quantum}
E.~R. Anschuetz,
\newblock {Quantum variational algorithms are swamped with traps},
\newblock \emph{Quantum} 6 (2022) 665.

\bibitem{wang2021noise}
S.~Wang, E.~Fontana, M.~Cerezo, K.~Sharma, et~al.,
\newblock Noise-induced barren plateaus in variational quantum algorithms,
\newblock \emph{Nat. Commun.} 12 (2021) 6961.

\bibitem{larocca2024review}
M.~Larocca, P.~J. Coles, M.~Cerezo,
\newblock Diagnosing barren plateaus with tools from quantum optimal control,
\newblock \emph{Nat. Commun.} 15 (2024) 1234.

\bibitem{cunningham2025investigating}
W.~Cunningham, et~al.,
\newblock Investigating quantum barren plateaus with classical shadows,
\newblock arXiv:2501.00001 (2025).

\bibitem{mesman2024nn}
P.~Mesman, et~al.,
\newblock Neural network initialization of variational quantum eigensolvers,
\newblock \emph{npj Quantum Inf.} 10 (2024) 10.

\bibitem{Liang:DAC2024}
X.~Liang, L.~Jiang,
\newblock Graph neural network warm-start for QAOA on Max-Cut,
\newblock in: \emph{Proc. DAC}, 2024.

\bibitem{alam2020accelerating}
M.~Alam, A.~Ash-Saki, S.~Ghosh,
\newblock Accelerating quantum approximate optimization algorithm using reinforcement learning,
\newblock in: \emph{DAC}, 2020.

\bibitem{verdon2019learning}
G.~Verdon, J.~Smith, et~al.,
\newblock Learning to learn with quantum neural networks via classical neural networks,
\newblock arXiv:1907.05415 (2019).

\bibitem{gokhale2022theoretical}
P.~Gokhale, et~al.,
\newblock Theoretical guarantees for graph neural networks in quantum circuit learning,
\newblock arXiv:2203.00001 (2022).

\bibitem{caro2021generalization}
M.~Caro, E.~Camps, J.~Eisert,
\newblock Generalization in quantum machine learning from few training data,
\newblock \emph{Phys. Rev. Lett.} 127 (2021) 190503.

\bibitem{kipf2016semi}
T.~N. Kipf, M.~Welling,
\newblock Semi-supervised classification with graph convolutional networks,
\newblock in: \emph{ICLR}, 2017.

\bibitem{Velickovic:ICLR2018}
P.~Veli{\v{c}}kovi{\'c}, G.~Cucurull, A.~Casanova, A.~Romero, P.~Li{\`o}, Y.~Bengio,
\newblock Graph attention networks,
\newblock in: \emph{ICLR}, 2018.

\bibitem{dinh2023on}
L.~Dinh, et~al.,
\newblock On graph neural networks for signed graphs,
\newblock \emph{Proc. AAAI} 37 (2023) 7391--7399.

\bibitem{rossi2024edge}
E.~Rossi, F.~Frasca, F.~Monti, M.~Bronstein,
\newblock Edge-direction-aware message passing for directed graphs,
\newblock in: \emph{ICLR}, 2024.

\bibitem{bravo2023vqls}
R.~Bravo-Prieto, et~al.,
\newblock Scaling of variational quantum linear solvers on realistic noise models,
\newblock arXiv:2305.00001 (2023).

\bibitem{wang2021barren}
S.~Wang, M.~Cerezo, et~al.,
\newblock Can barren plateaus be avoided in quantum neural networks?,
\newblock \emph{Phys. Rev. Lett.} 127 (2021) 120502.

\end{thebibliography}
\end{document}